\documentclass[prl,twocolumn,superscriptaddress,showpacs]{revtex4}

\usepackage{graphicx}

\begin{document}
\title{Comment on ``The Quantum State of a Propagating Laser Field''}

\author{Terry Rudolph}
\affiliation{Bell Labs, 600-700 Mountain Ave., Murray Hill, NJ 07974, U.S.A.}
\author{Barry C.\ Sanders}
\affiliation{Department of Physics, Macquarie University, Sydney,
        New South Wales 2109, Australia}
\affiliation{Quantum Entanglement Project, ICORP, JST,
    Ginzton Laboatory, Stanford University,
    California 94305-4085}

\date{\today}

\begin{abstract}
We comment on the theoretical quantum state of a propagating
laser field proposed by van Enk and Fuchs
[quant-ph/0104036, quant-ph/0111157]
and clarify that the multimode description of the propagating
laser field does not modify our analysis of continuous variable
quantum teleportation [quant-ph/0103147].
Furthermore we point out that
the ``complete measurements'' discussed by van Enk and Fuchs
have not been achieved by existing technology and may not be
possible even in principle.
\end{abstract}
\maketitle

In critiquing our assertion that continuous variable quantum
teleportation (CVQT) cannot be demonstrated with a conventional
laser in a linear optical system even with ideal photodetectors and
perfect Fock state sources allowed~\cite{Rud01},
van Enk and Fuchs (vEF)~\cite{Enk01a} analyze
a multimode propagating laser field (MPLF) and claim that a
conventional laser suffices.  They agree that the intracavity
field is correctly described by a mixed state that is diagonal in
the Fock state basis, and consequently the coupled--mode analysis
they employ yields the MPLF outside the cavity to be a mixed
state. However, they go on to demonstrate the attractiveness of
expressing the MPLF as an ensemble of multimode coherent states
in preference to the many other possible ensembles, with the
quantum de Finetti theorem employed in support of their
intuitive preference for this particular decomposition of the MPLF
density matrix. Specifically, the quantum de Finetti theorem can be
used to show that the coherent state ensemble is the only one
which conforms with their particular intuition.  Such reasoning is
clearly open to question however. For example, the mixed state
MPLF admits a decomposition that is diagonal in the multimode
Fock basis, with each pure state in this ensemble exhibiting
manifest photon number entanglement between the modes. This Fock
basis ensemble is the only one which requires no violation of
conservation of energy to prepare: it too is unique with respect
to an intuition that many physicists share.

Although we acknowledge the attraction of the coherent state
decomposition as a convenient tool for simplifying calculations,
designing experiments, and as motivation to strive for vEF's
``complete measurement'', vEF's approach does not establish a
necessary condition for the coherent state decomposition being
privileged. In effect they establish an argument which allows for
a {\it sufficient} understanding of the experiment; however, we
believe that claims to demonstrate CVQT must establish that the
experiment is {\it necessarily} interpreted as such. Our view is
that in quantum information processing the use of preferred
ensembles can be justified only in scenarios involving
retrodiction of preparation procedures, and this retrodictive
rationale has no relevance to the topic of CVQT using lasers.
%
%For example, the mixed state MPLF admits a decomposition that is
%diagonal in the multimode Fock basis, with each pure state in
%this ensemble exhibiting manifest photon number entanglement
%between the modes.

As a justification for favouring coherent states, vEF discuss
``complete measurements'' that purportedly could be employed to
project the MPLF into a true coherent state with known phase; the
claim is that
 inter-modal phase correlation allows such a
``complete measurement'' on one mode of the MPLF to reduce the
mixed state to a pure multimode coherent state.  Despite common
misconceptions that phase or phase-sensitive measurements have
been performed or are possible, no experiment has either been
performed or even proposed \emph{in principle} which would yield
an observation that depends in any way upon the relative phase of
a superposition of Fock number states within a given mode. The
phase of the laser field is just such a phase (it affects the
relative phase between superpositions of number states), and the
so--called complete measurement would therefore have to be of a
completely unconventional type. Measuring all orders of the
standard photon correlation functions \cite{Gla63} cannot yield
information about this phase.

We do not, however, accept vEF's premise that some kind of
unconventional measurement can in fact measure the phase in a way
that necessitates updating the mixed--state description of the
MPLF to a pure coherent state. Such a measurement would violate
the energy conservation principle. As with any conserved quantity,
the description of such a measurement may be constructed by
invoking an ``effective classical field'' approximation, such as
employing the classical field description of the local oscillator
in homodyne detection, but a rigorous description of the
measurement, which does not invoke the mean field approximation
and does not yield a coherent superposition of energy states in a
specific mode, is always possible and is certainly as valid.
Moreover, unless some measurement can be
devised, or some operational procedure developed, for which the
vEF preferred ensemble may be subjected to empirical tests,
we see no reason to prefer their ensemble.

On a separate point,
a key criticism by vEF is our alleged claim that there was no
entanglement in the experiment~\cite{Fur98}, hence no CVQT. This
was not our claim. Rather we demonstrated that the two--mode
squeezed field is not entangled (about which vEF agreed) to
demonstrate the danger of the preferred ensemble fallacy (PEF).
In Fig.~1 we give yet another simple example of the danger
inherent in analyzing CVQT using preferred ensembles.

\begin{figure}
\includegraphics*[width=3in,keepaspectratio]{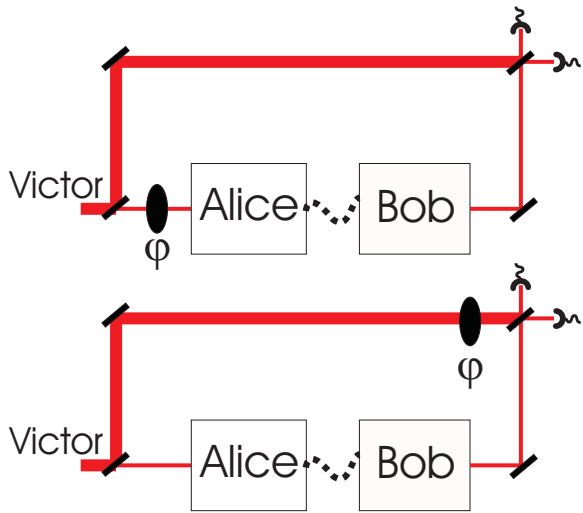}
\caption{Two diagrams indicating another simple danger with
using the preferred ensemble of coherent states when discussing an
experiment. We consider the case where Victor has an independent
laser and chooses to verify CVQT by injecting his laser beam into
a Mach--Zehnder interferometer, with CVQT performed in one
channel of the interferometer and the other ``reference'' channel
being retained by him. In the upper diagram Victor uses a phase
shifter with a number of different settings
$\phi_0,\phi_1,\phi_2,\ldots$, which he varies in order to test
Alice and Bob's teleportation. A proponent of coherent states
would assert that the teleportation is being tested by Victor
giving Alice a series of different, non-orthogonal coherent
states $|\alpha_0\rangle,|\alpha_1\rangle,|\alpha_2\rangle,\ldots$
to be teleported.
However, an operationally  equivalent verification can be obtained
by Victor placing the phase shifter in his reference beam {\it
after} the teleportation has allegedly taken place, as indicated
in the lower diagram. Examples such as this one and others in
\cite{Rud01} illustrate the ambiguity of describing ``what really
happened'' in any given experiment by using a preferred ensemble;
we therefore feel compelled to reject such decomposition-dependent 
descriptions.}
\end{figure}

In summary our objection to the claim that CVQT has been
achieved~\cite{Fur98} is that its apparent success requires a
preferred ensemble.  As the quantum teleportation protocol refers
to initial states and their subsequent evolution, the
multiplicity of differing descriptions for specific data are
equally valid.  A reconciliation between our objection and the
claim of successful CVQT is presumably possible, however, if the
CVQT protocol is reconstructed in terms of an initial mixed
state. The challenge is then to agree upon a set of genuine
operational criteria for ``unconditional'' CVQT.

\begin{acknowledgments}
This work has been partially supported by an Australian Research Council Large Grant.
BCS appreciates valuable discussions with S.\ D.\ Bartlett. TR has
enjoyed stimulating discussions with S.\ J.\ van Enk and Ch.\ Fuchs
who have been kind enough to provide us with advance copies of their
eprints.
\end{acknowledgments}

\end{document}